\documentclass[letterpaper,10pt,conference]{ieeeconf}
\IEEEoverridecommandlockouts
\usepackage{amsmath,amssymb,mathtools}

\usepackage{amsthm}
\usepackage{graphicx,booktabs,cite}
\usepackage{balance}
\usepackage{float}
\let\labelindent\relax
\usepackage{enumitem}
\floatstyle{ruled}
\newfloat{algorithm}{tbp}{loa}
\floatname{algorithm}{Algorithm}
\usepackage[hidelinks]{hyperref}
\hypersetup{pdftitle={SafePG with Hard Intersample Safety Guarantees},
 pdfauthor={Vipul K. Sharma, S. Sivaranjani}}
\newtheorem{theorem}{Theorem}[section]
\newtheorem{lemma}[theorem]{Lemma}
\newtheorem{proposition}[theorem]{Proposition}

\newtheorem{assumption}[theorem]{Assumption}
\theoremstyle{remark}
\newtheorem{remark}[theorem]{Remark}

\makeatletter
\g@addto@macro\appendices{%
}
\makeatother
\newcommand{\R}{\mathbb R}
\newcommand{\E}{\mathbb E}
\newcommand{\U}{\mathcal U}
\newcommand{\Ssafe}{\mathcal S}
\newcommand{\Cdel}{\mathcal C_\delta}
\newcommand{\pdel}{\pi_\theta^\delta}
\DeclarePairedDelimiter{\norm}{\lVert}{\rVert}

\newcommand{\Geom}{\operatorname{Geom}}
\newcommand{\diam}{\operatorname{diam}}

\usepackage{algpseudocode}

\usepackage{amsmath,amssymb,graphicx}
\usepackage{xcolor}

\newcounter{vipcounter}

\title{\LARGE\bf Safe Reinforcement Learning with Hard Constraints:\\ Policy Gradient Methods with Intersample Guarantees}
\author{Vipul K. Sharma and S. Sivaranjani%
\thanks{Vipul K. Sharma and S. Sivaranjani are with the Edwardson School of
Industrial Engineering, Purdue University, West Lafayette, IN 47907, USA.
\texttt{\{sharm697,sseetha\}@purdue.edu}.}%
}
\usepackage{amsmath,amssymb,graphicx}

\begin{document}
\maketitle
\thispagestyle{empty}
\pagestyle{empty}
\begin{abstract}
We develop a policy-gradient (PG) reinforcement learning (RL) framework with hard inter-sample control barrier function (CBF)-based safety constraints for nonlinear sampled-data systems. We first design a class of stochastic policies with support truncated to sampling-period-dependent safe control sets, that guarantee safety-by-construction at sampling instants. We then propose a model-free PG algorithm that searches over the class of safe-by-design policies and establish sufficient conditions on sampling-period such that any action sampled from safe control sets remains safe throughout the sampling interval, providing inter-sample guarantees. We further establish finite-time convergence and sample complexity guarantees for the proposed PG algorithm and characterize its global optimality under gradient dominance. Finally, we illustrate inter-sample safety guarantees on a safe quadrotor navigation environment.


\end{abstract}
\section{Introduction}\label{sec:introduction}
Reinforcement learning (RL) enables learning control policies by
interacting with an unknown environment. However, in safety-critical
applications such as autonomous driving and robotics, hard safety constraints must be satisfied throughout learning and
deployment. This requirement is particularly challenging in model-free
RL, where stochastic exploration is used to improve the policy and
can lead to constraint violations before an effective controller is
learned. For physical systems, these constraints must hold along the
continuous-time trajectories generated by the controller, including
between successive control updates.

In the RL literature, safety is typically addressed through
constrained Markov decision processes, primal-dual methods, or
penalty-based approaches~\cite{achiam2017constrained}. These methods
regulate constraint violations through expected or cumulative costs,
but do not, in general, guarantee pointwise satisfaction of hard
constraints during exploration. In learning-based control, hard
constraints are commonly enforced through control barrier function
(CBF)-based safety filters, where the actions of a potentially unsafe
RL policy are projected onto a safe control
set~\cite{ames2019cbf,cheng2019end,cohen2023safe,hsu2024filter,yang2026cbf}. 
Although such filters can ensure safety under suitable conditions,
the projection changes the policy applied to the system. Consequently, the
convergence and optimality guarantees of the underlying unconstrained
RL algorithm do not directly extend to the filtered
policy~\cite{suttle2024sampling}.

Another line of work incorporates hard constraints into the policy
class itself and optimizes directly over these safe-by-design policies. Accounting
for the constrained exploration distribution in the objective and its
gradient allows convergence to be analyzed for the policies actually
used during learning and deployment.
For example, the safe RL framework in~\cite{suttle2024sampling} establishes
convergence while satisfying hard constraints during learning and
deployment. More recently, SafePG~\cite{sharma2026safepg} and
Safe-PG-LQR~\cite{sharma2026safepglqr} develop safe policy-gradient
methods with optimality guarantees for nonlinear control and
constrained LQR, respectively. These results build on policy
optimization over structured classes, for which convergence and
optimality can be characterized through properties such as gradient
dominance~\cite{fazel2018global,sharma2025globally}.

These safe-by-design frameworks only consider discrete-time dynamics.
In most practical control applications, plants evolve with continuous-time dynamics, and are controlled using digital hardware, thus yielding sampled-data dynamics. In such cases, safe learning approaches developed for discrete-time dynamics, enforcing
safety constraints only at sampling instances, can result in potential safety violations when deployed in sampled-data systems. 
Sampled-data CBF methods address this by ensuring intersample safety
through an additional barrier margin at each sampling
instant~\cite{singletary2020control,breeden2022sampled,bahati2023samplehold},
including settings with uncertainty in the barrier
dynamics~\cite{begzadic2026sampled}. Other approaches safeguard RL
controllers in continuous time through reachability-based action
projection~\cite{kochdumper2023projection} and sampled-data funnel
control~\cite{lanza2024sampled}. 

In this paper, we develop a safe policy optimization framework for
nonlinear sampled-data systems with intersample safety, convergence,
and optimality guarantees. We consider a stochastic policy class that facilitates exploration, truncate its support to enforce CBF-type safety constraints by design, and optimize over this safe-by-design policy class. 
Our focus is to incorporate intersample
safety into this stochastic policy class and analyze convergence for
the resulting RL objective. Our main contributions are as follows:
\begin{itemize}
\item First, we construct a sampling-aware safe control set using a
lower bound on the CBF expression and a bound on its variation over
each sampling interval. We establish a sufficient condition on the
sampling period for feasibility and show that selecting inputs from
this set preserves state and actuator constraints throughout each
interval. Thus, truncating the support of a base stochastic policy to this
set gives a safe-by-design policy class with intersample guarantees
during exploration and deployment.
\item Second, we derive the score-function and policy-gradient
expressions for this truncated policy class. We construct an unbiased model-free policy
gradient estimator, establish smoothness and variance
bounds, and characterize the expected number of plant interactions
required to obtain a gradient estimate.
\item Third, we develop a sampling-aware safe policy gradient (PG)
algorithm, \texttt{SASPG}, and establish finite-time convergence and
sample-complexity guarantees under weak gradient dominance. These
guarantees characterize optimality while preserving intersample safety in nonlinear sample-data systems. Finally, we demonstrate the framework on a safe quadrotor navigation problem.
\end{itemize}
\subsection*{Organization}
We define the sampled-data dynamical system and associated Markov Decision Process, and formulate the safe RL problem Section~\ref{sec:problem}. Then, we establish sufficient conditions on the sampling period for intersample safety guarantees in Section~\ref{sec:safety}.
Using these conditions, we define a class of safe-by-design
policies, formulate the safe RL problem with intersample safety guarantees and propose the safe sampling-aware PG algorithm in Section~\ref{sec:policy}. Furthermore, we provide convergence, optimality, and
sample complexity guarantees for the proposed PG algorithm in Section~\ref{sec:convergence}, and implement it on a safe quadrotor navigation environment in Section~\ref{sec:safe_quadrotor_experiment}. Finally, we provide proofs our theoretical results in Appendix~\ref{app:proofs}.

\subsection*{Notation}\label{subsec:notation}
We denote the real numbers and $n$-dimensional real vectors by
$\R$ and $\R^n$, respectively. For a matrix $A$, $A^\top$
denotes its transpose, and $I_d$ is the $d\times d$ identity matrix.
The notation $\norm{\cdot}$ denotes the Euclidean norm for vectors
and its induced norm for matrices. We write $\nabla_\theta$
and $D_\theta$ for the gradient and Jacobian with respect to $\theta$.
We use $\mathbf1_A$ for the indicator of a set $A$,
$\lambda_m$ for $m$-dimensional Lebesgue measure, and
$\diam(A):=\sup_{x,y\in A}\norm{x-y}$ for its diameter.
We use the notation $\E$ to represent expectation of the appropriate random variables. We use the abbreviation a.s. to mean `almost surely', and for $z \in(0,1)$, use $\Geom(z)$ to denote geometric
distribution on $\{0,1,\ldots\}$ with probability
$z(1-z)^k$ at $k$. Finally we use $\operatorname{clip}(x,a,b)$ as the clip operator, limiting variable $w\in \R^w$ to $[-a,b]$, component-wise.

\section{Problem Formulation}\label{sec:problem}
\subsection{System and Sampled MDP}
We consider a control-affine nonlinear system
\begin{equation}\label{eq:plant}
 \dot x=f(x)+g(x)u,
\end{equation}
where $x\in\mathcal X=\R^n$ and $u\in\U=\R^m$ are the state and input of the system, and functions $f,g$ are locally Lipschitz. 

In practical implementations, control policies are executed on digital
hardware, such as microcontrollers, which update the control input at
a finite rate determined by sensing, computation, and actuation
requirements. For a sampling period $\delta>0$, let $t_k=k\delta$
and $x_k=x(t_k)$. Under a zero-order hold, the control input is held constant between
successive sampling instants:
\begin{equation}\label{eq:zoh}
u(t)=u_k,\qquad t\in[t_k,t_{k+1}).
\end{equation}
For a constant input $u$, suppose $x(s;x,u)$ is the solution
of~\eqref{eq:plant} at time $s$ with initial state $x$, for all $s\in[0,\delta]$. Then, we write the sampled dynamics
as
\begin{equation}\label{eq:sampled-dynamics}
x_{k+1}=F_\delta(x_k,u_k),\qquad
F_\delta(x,u):=x(\delta;x,u).
\end{equation}

This zero-order-hold dynamics induces the sampled transition kernel $P_\delta$, for every Borel set $A\subseteq\mathcal X$, defined as
\begin{equation}\label{eq:transition-kernel}
 P_\delta(A\mid x,u)=\mathbf1_A(F_\delta(x,u)).
\end{equation}
To explore the environment and learn from sampled interactions,
we use a stochastic feedback policy $\pi(\cdot\mid x)$ that
samples an input $u_k\sim\pi(\cdot\mid x_k)$ at each update.
Let $r:\mathcal X\times\U\to\R$ be a measurable per-step reward,
satisfying the bound below, and choose a discount factor
$\gamma\in(0,1)$ independently of $\delta$:
\begin{equation}\label{eq:rewardbound}
 |r(x,u)|\le\bar r,\qquad (x,u)\in\mathcal X\times\U.
\end{equation}
We now define the following discounted MDP, based on the sampled-data system, as
\begin{equation}\label{eq:mdp}
 \mathsf M_\delta:=(\mathcal X,\U,P_\delta,r,\gamma),
\end{equation}
with a fixed initial-state distribution $x_0\sim\mu_0$. We use this MDP to formulate the safe RL problem, where safety is described using control barrier function (CBF) based forward invariance constraints as follows.

\subsection{Safety}
Our notion of safety requires the state of~\eqref{eq:plant} to
remain in a prescribed subset $\Ssafe\subseteq\mathcal X$,
called the safe set, described as the
zero-superlevel set of a continuously differentiable function
$h:\mathcal X\to\R$ as follows:
\begin{equation}\label{eq:safe-set}
 \Ssafe:=\{x\in\mathcal X:h(x)\ge0\}.
\end{equation}
The set $\Ssafe$ is forward invariant if
every trajectory with $x(0)=x_0\in\Ssafe$ satisfies
$x(t)\in\Ssafe$ for all $t>0$.
From~\eqref{eq:safe-set}, this is equivalent to preserving
$h(x(t))\ge0$ whenever $h(x_0)\ge0$.
A sufficient condition ensuring such a forward invariance property is the following CBF-based condition~\cite{ames2017qp,ames2019cbf}
\begin{equation}\label{eq:first-order-invariance}
 \dot h(x(t))+\alpha_1(h(x(t)))\ge0
 \quad \forall t\geq 0,
\end{equation}
where $\alpha_1$ is a locally Lipschitz extended class-$\mathcal K$
function. 
%

For system \eqref{eq:plant}, $\dot h=L_fh+L_gh\,u$, with Lie derivatives
$L_fh=\nabla h^\top f$ and $L_gh=\nabla h^\top g$, we define
\begin{equation}\label{eq:psi}
 \psi(x,u):=L_fh(x)+L_gh(x)u+\alpha_1(h(x)).
\end{equation}
The barrier inequality~\eqref{eq:first-order-invariance} can therefore
be written as $\psi(x,u)\ge0$, giving a state-dependent constraint
on the control input. Before we design the set of safe control inputs, we also consider the actuator limits
$u\in\U_{\mathrm{act}}$ as
\begin{equation}\label{eq:actuator-limits}
 \U_{\mathrm{act}}:=\prod_{i=1}^m[\underline u_i,\overline u_i]
 \subset\R^m,
\end{equation}
where $-\infty<\underline u_i<\overline u_i<\infty$ and $D_{\mathrm{act}}:=\diam(\U_{\mathrm{act}})$.

We call $h$ a control barrier function (CBF) for
system~\eqref{eq:plant} with input constraints $u\in\U_{\mathrm{act}}$
if there exists a locally Lipschitz extended class-$\mathcal K$
function $\alpha_1$ such that
\begin{equation}\label{eq:first-order-cbf}
 \sup_{u\in\U_{\mathrm{act}}}\psi(x,u)\ge0,
 \qquad x\in\Ssafe.
\end{equation}
This condition ensures that, at every state in $\Ssafe$, at least
one input satisfies both the CBF and actuator constraints.
These inputs form the safe control set
\begin{equation}\label{eq:cbf-actions}
 K_{\mathrm{cbf}}(x)
 :=\{u\in\U_{\mathrm{act}}:\psi(x,u)\ge0\}.
\end{equation}
%
\begin{remark}
When designing the safe control set, the barrier inequality must
explicitly constrain the input $u$ through time derivatives of $h$. When $u$ does not appear explicitly in
$\dot h$, we consider high-order CBFs to obtain an input-dependent inequality
using higher derivatives of $h$~\cite{xiao2022hocbf}.
\end{remark}

\subsection{Safe RL for Sampled-Data Systems}
Consider the sampled MDP $\mathsf M_\delta$ in~\eqref{eq:mdp},
and let $\Pi$ denote the class of measurable stochastic feedback
policies on $\U$. Each $\pi\in\Pi$ induces a trajectory distribution
through $x_0\sim\mu_0$, $u_k\sim\pi(\cdot\mid x_k)$, and
$x_{k+1}\sim P_\delta(\cdot\mid x_k,u_k)$. We write $\E_\pi$
for expectation under this distribution. Our goal is to find a policy in this class
that maximizes the expected discounted return while satisfying
the prescribed safety constraints as
\begin{equation}\label{eq:safe-rl-problem}
\begin{aligned}
 \sup_{\pi\in\Pi}\quad&\E_\pi\!\left[
       \sum_{k=0}^\infty\gamma^k r(x_k,u_k)\right]\\
 \text{s.t.}\quad
   &x_k\in\Ssafe,\quad u_k\in K_{\mathrm{cbf}}(x_k)
       \quad\text{a.s.},\\
   &x_{k+1}\sim P_\delta(\cdot\mid x_k,u_k),\quad k\ge0.
\end{aligned}
\end{equation}
Here, $x_k\in\Ssafe$ requires the state to remain safe at each
sampling instant, and $u_k\in K_{\mathrm{cbf}}(x_k)$ requires
the selected input to satisfy both the CBF and actuator constraints.

Safe policy-gradient methods, such as
\cite{suttle2024sampling,sharma2026safepg,sharma2026safepglqr}, address
hard-constrained policy optimization for discrete-time systems by
restricting exploration to a safe control set.
However, in such safe RL formulations, the safety guarantees
only hold at update times, and do not, in
general, establish safety between control updates, that is, \textit{intersample safety}.
In particular, $x_k,x_{k+1}\in\Ssafe$ does not imply that
$x(k\delta+s)\in\Ssafe$ for every $s\in[0,\delta]$ under the zero-order-hold input. We next address this gap by first constructing a sampling-period-dependent formulation of $K_{\mathrm{cbf}}(x)$ and establish sufficient conditions for intersample safety and feasibility. 

\section{Intersample Safety and Feasibility}\label{sec:safety}
Under the zero-order hold~\eqref{eq:zoh}, barrier expression \eqref{eq:psi}
can change while the input remains constant. We account for this
variation by introducing an additional margin at each update.

%
\begin{assumption}\label{ass:safety}
Let $\delta_0>0$ and $M_\psi>0$ be known constants, and let
$\underline\psi:\mathcal X\times\U_{\mathrm{act}}\to\R$ be a known
jointly measurable function, continuous in $u$.
For every $x\in\Ssafe$ and constant $u\in\U_{\mathrm{act}}$,
we have $\psi(x,u)\ge\underline\psi(x,u)$. Furthermore, the solution
$x(s;x,u)$ exists on $[0,\delta_0]$ and satisfies
$\psi(x(s;x,u),u)\ge\psi(x,u)-M_\psi s$
for all $s\in[0,\delta_0]$.
\end{assumption}

This lower bound can account for model uncertainty in unknown environment. For example,
an estimate $\widehat\psi$ satisfying
$|\widehat\psi(x,u)-\psi(x,u)|\le\varepsilon_\psi(x,u)$ gives
$\underline\psi=\widehat\psi-\varepsilon_\psi$.
Such bounds can be obtained from validated model or estimation
errors, including learning-based sampled-data CBF
constructions~\cite{begzadic2026sampled}.

Similarly, if $\psi(\cdot,u)$ has a uniform Lipschitz constant
$L_\psi$ and $\norm{x(s;x,u)-x}\le Vs$, then
$M_\psi\ge L_\psi V$ is a sufficient variation bound. These bounds must hold on a region containing all trajectories
from $\Ssafe$ under constant inputs in $\U_{\mathrm{act}}$
or $0\le s\le\delta_0$.
Based on these lower bounds, we characterize the sampling-aware safe
control set, for $0<\delta\le\delta_0$, as
\begin{equation}\label{eq:SASafeSet}
\Cdel(x):=\{u\in\U_{\mathrm{act}}:
                 \underline\psi(x,u)\ge M_\psi\delta\}.
\end{equation}
Since $\psi\geq \underline\psi$, we have
$\Cdel(x)\subseteq K_{\mathrm{cbf}}(x)\subseteq\U_{\mathrm{act}}$.
The margin $M_\psi\delta$ compensates for the possible decrease
in the barrier expression before the next update.

\begin{theorem}[Intersample Safety]\label{thm:safety}
Suppose Assumption~\ref{ass:safety} holds and $0<\delta\le\delta_0$.
If $x_0\in\Ssafe$ and $u_k\in\Cdel(x_k)$ at every update, then
the trajectory under~\eqref{eq:zoh} satisfies
$x(t)\in\Ssafe$ and $u(t)\in\U_{\mathrm{act}}$ for all $t\ge0$.
\end{theorem}
\noindent\emph{Proof:} See Appendix~\ref{app:proofs}, item~(\ref{app:proof-safety}).\par

To select a safe input at every update, we require \textit{feasibility} of the set $\Cdel(x)$, that is {$\Cdel(x)\ne\emptyset$} for every $x\in\Ssafe$. Let
$m_{\mathrm{act}}:=\inf_{x\in\Ssafe}
\max_{u\in\U_{\mathrm{act}}}\underline\psi(x,u)$
denote a uniform lower bound on the largest barrier margin
attainable within the actuator limits. The following proposition
gives a sufficient condition on $\delta$ for this
\textit{feasibility} requirement to hold.

\begin{proposition}\label{prop:period}
Suppose Assumption~\ref{ass:safety} holds. If $m_{\mathrm{act}}>0$
and the sampling period satisfies
\begin{equation}\label{eq:period}
 0<\delta\le\delta_0,\qquad \delta<m_{\mathrm{act}}/M_\psi,
\end{equation}
then $\Cdel(x)\ne\emptyset$ for every $x\in\Ssafe$.
\end{proposition}
\noindent\emph{Proof:} See Appendix~\ref{app:proofs}, item~(\ref{app:proof-period}).\par

\section{Intersample Safe Policy Gradient}\label{sec:policy}
We now use the sampling-aware safe control set \eqref{eq:SASafeSet} to construct a class
of stochastic policies that preserves intersample safety during exploration, and formulate a Safe RL problem that directly searches over this class of policies.

\subsection{Intersample Safe-by-Design Policies}
With Theorem~\ref{thm:safety}, we know that inputs selected from $\Cdel(x)$
preserve safety throughout each sampling interval. We incorporate
this guarantee into stochastic policies by truncating their support to
$\Cdel(x)$. To define a density on this set, we require
$\lambda_m(\Cdel(x))>0$. Under the conditions of
Proposition~\ref{prop:period}, the strict barrier margin and
continuity of $\underline\psi(x,\cdot)$ ensure this property
at every $x\in\Ssafe$.

For a jointly measurable base-policy density $\pi_\theta(u\mid x)$
on $\U=\R^m$, $\theta\in\R^p$, define the safe policy by
\begin{equation}\label{eq:policy}
 \pdel(u\mid x)
 =\frac{\pi_\theta(u\mid x)\mathbf1_{\Cdel(x)}(u)}
         {Z_\theta^\delta(x)},
\end{equation}
where $Z_\theta^\delta(x):=\int_{\Cdel(x)}\pi_\theta(v\mid x)\,dv$
is the probability that the base policy selects a safe input.
We require $Z_\theta^\delta(x)>0$ for every
$(\theta,x)\in\R^p\times\Ssafe$.
%
%

We take the parameter space to be $\Theta=\R^p$ and
restrict~\eqref{eq:safe-rl-problem} to the safe-by-design policy
class $\Pi_\delta:=\{\pi_\theta^\delta:\theta\in\R^p\}$.
For a fixed sampling period satisfying~\eqref{eq:period}, the
trajectory law is determined by
$x_0\sim\mu_0$, $u_k\sim\pdel(\cdot\mid x_k)$, and
$x_{k+1}\sim P_\delta(\cdot\mid x_k,u_k)$.
We write $\E_\theta^\delta$ for expectation under this law and
define the return of a policy in $\Pi_\delta$ as
\begin{equation}\label{eq:rl-return}
 J_\delta(\theta)
 :=\E_\theta^\delta\!\left[
       \sum_{k=0}^\infty\gamma^k r(x_k,u_k)\right].
\end{equation}
The reformulated safe RL problem is therefore
\begin{equation}\label{eq:intersample-safe-rl}
 \sup_{\theta\in\R^p}J_\delta(\theta)=:J_\delta^*.
\end{equation}
The state and actuator constraints are enforced by the policy support
in~\eqref{eq:policy}, so~\eqref{eq:intersample-safe-rl} searches
directly over safe policies. 
Finally, the reward bound~\eqref{eq:rewardbound} gives
\begin{equation}\label{eq:returnbound}
 |J_\delta(\theta)|\le\frac{\bar r}{1-\gamma}.
\end{equation}
Thus, the reformulated safe RL problem is well-defined, and we can now develop policy gradient algorithms solving this problem.
\subsection{Safe Policy Gradient}
We first state the conditions needed to differentiate the objective
and bound its gradient. Suppose the base-policy score is defined as
$b_\theta(x,u):=\nabla_\theta\log\pi_\theta(u\mid x)$ and the
truncated-policy score as
$s_\theta^\delta(x,u):=\nabla_\theta\log\pdel(u\mid x)$.
For a finite rollout $\tau_n=(x_0,u_0,\ldots,u_{n-1})$,
the states follow $x_{k+1}=F_\delta(x_k,u_k)$ and the trajectory
density is $p_{\theta,n}^\delta(\tau_n)
=\prod_{k=0}^{n-1}\pdel(u_k\mid x_k)$ with respect to the reference measure
$\nu_n(d\tau_n):=\mu_0(dx_0)\,du_0\cdots du_{n-1}$,
which combines the initial-state distribution with Lebesgue
measure over the sampled actions.

%
\begin{assumption}[Policy Regularity]\label{ass:policy}
The base-policy class satisfies the following conditions.
\begin{enumerate}[label=(\roman*),ref=\roman*]
\item For every $x\in\Ssafe$ and $u\in\Cdel(x)$,
$\pi_\theta(u\mid x)>0$ for all $\theta\in\R^p$, and
$\theta\mapsto\pi_\theta(u\mid x)$ is twice continuously
differentiable.
\item\label{asum-item-integrability}
For every compact parameter set $B\subset\R^p$, state
$x\in\Ssafe$, and finite horizon $n\ge1$, the norms of the first two
parameter derivatives of $\pi_\theta(\cdot\mid x)$ and
$p_{\theta,n}^\delta$ admit dominating functions uniform over
$\theta\in B$. These functions are integrable with respect to
$du$ on $\Cdel(x)$ and $\nu_n$, respectively.
\item\label{asum-item-bound}
There are finite constants $M_{s,2},M_{D_s}\ge0$ such that,
for every $(\theta,x)\in\R^p\times\Ssafe$ and
$u\sim\pdel(\cdot\mid x)$,
$\E[\norm{s_\theta^\delta(x,u)}^2]\le M_{s,2}$ and
$\E[\norm{D_\theta s_\theta^\delta(x,u)}]\le M_{D_s}$.
\end{enumerate}
\end{assumption}
Assumption~\ref{ass:policy}(\ref{asum-item-integrability}) permits
differentiation of policy and trajectory integrals, while
Assumption~\ref{ass:policy}(\ref{asum-item-bound}) bounds the score
and its sensitivity to $\theta$.
Since $\Cdel(x)$ is independent of $\theta$, the truncated-policy
score satisfies
\begin{equation}\label{eq:score}
 s_\theta^\delta(x,u)
 =b_\theta(x,u)
  -\E_{v\sim\pdel(\cdot\mid x)}[b_\theta(x,v)].
\end{equation}
In particular, its conditional mean is zero.
Further, for $x\in\Ssafe$ and $u\in\Cdel(x)$, we define the action-value
function
\begin{equation}
    Q_\theta^\delta(x,u)
 :=\E_\theta^{\delta,x,u}
 \left[\sum_{k=0}^{\infty}\gamma^k r(x_k,u_k)\right],
\end{equation}
where the rollout starts at $(x_0,u_0)=(x,u)$ and follows
$\pdel$ thereafter. We further define the following distribution 
\begin{equation}
    d_\theta^\delta:=(1-\gamma)\sum_{k=0}^{\infty}
       \gamma^k\mathcal L_\theta^\delta(x_k),
\end{equation}
where $\mathcal L_\theta^\delta(x_k)$ is the state distribution induced
by $\mu_0$, $\pdel$, and $P_\delta$.
Then, the policy-gradient theorem~\cite{sutton1999policy} gives
\begin{equation}\label{eq:pg}
 \nabla J_\delta(\theta)
 =\frac{1}{1-\gamma}
   \E_{x\sim d_\theta^\delta,u\sim\pdel(\cdot\mid x)}\!\left[Q_\theta^\delta(x,u)
             s_\theta^\delta(x,u)\right].
\end{equation}
We now present the following smoothness result.
\begin{lemma}\label{lem:score}
Suppose Assumptions~\ref{ass:safety} and~\ref{ass:policy} hold and
$\delta$ satisfies~\eqref{eq:period}. Then
$\norm{\nabla J_\delta(\theta')-\nabla J_\delta(\theta)}
\le L_\delta\norm{\theta'-\theta}$ for all
$\theta,\theta'\in\R^p$, where
\begin{equation}\label{eq:L}
 L_\delta=\bar r
 \left[\frac{2M_{s,2}(1+\gamma)}{(1-\gamma)^3}
              +\frac{M_{D_s}}{(1-\gamma)^2}\right].
\end{equation}
\end{lemma}
\noindent\emph{Proof:} See Appendix~\ref{app:proofs}, item~(\ref{app:proof-smoothness}).\par
This result implies that policy gradient does not change abruptly with parameter $\theta$, which is a desired condition for policy-gradient based search algorithms.
\subsection{Model-free Gradient Estimation}
We now construct an unbiased estimator of~\eqref{eq:pg} in two
stages. The first stage samples a state from $d_\theta^\delta$, and
the second estimates the return and policy score at that state.

In the first stage, we independently sample $x_0\sim\mu_0$ and
$K\sim\Geom(1-\gamma)$. Then, we follow $\pdel$ for
$K$ sampling intervals and set $x=x_K$. Since $K$ is independent
of the rollout, the terminal state $x$ follows distribution
$d_\theta^\delta$.

In the second stage, we draw $u,v$ conditionally independently
from $\pdel(\cdot\mid x)$ and define
$A_\theta(x,u,v):=b_\theta(x,u)-b_\theta(x,v)$.
The input $u$ starts the return rollout, while $v$ is an auxiliary
random variable used to center the base score. By~\eqref{eq:score},
$A_\theta$ is an unbiased estimate of the truncated-policy score
conditional on $(x,u)$, without differentiating the normalizer.

To estimate the return from $(x,u)$, we apply $u$ for one sampling
interval, then continue with $\pdel$ afterwards, and denote the corresponding samples
by $(\tilde x_h,\tilde u_h)$, with
$(\tilde x_0,\tilde u_0)=(x,u)$, and per-step reward
$r_h=r(\tilde x_h,\tilde u_h)$.
We define $q:=\sqrt\gamma$, independently sample
$H\sim\Geom(1-q)$, and define the return estimate as
\begin{equation}
    \widehat Q_\delta:=\sum_{h=0}^H q^h r_h.
\end{equation}
Given $(x,u)$, the policy draws used to generate the return
rollout are independent of the auxiliary sample $v$.
We sample $H$ independently of $v$ and of the randomness used
to generate this rollout, keeping $\theta$ fixed throughout.
The randomized horizon makes $\widehat Q_\delta$ an unbiased
estimate of $Q_\theta^\delta(x,u)$ by removing finite-horizon
truncation bias~\cite{suttle2024sampling}. Thus, we get the following gradient estimate
\begin{equation}\label{eq:ghat}
 \widehat g_\delta
 :=\frac{\widehat Q_\delta}{1-\gamma}A_\theta(x,u,v).
\end{equation}
We summarize the estimator construction steps in Algorithm~\ref{alg:safegrad},
\begin{algorithm}[htb]
\caption{\texttt{SafeGrad}: Safe Policy Gradient Estimator}
\label{alg:safegrad}
\begin{algorithmic}[1]
\Require $\theta$
    \State Draw $x_0\sim\mu_0$ and
    $K\sim\Geom(1-\gamma)$ independently
    \State Starting at $x_0$, follow $\pdel$ for
    $K$ sampling intervals and set $x\gets x_K$
    \State Draw $u,v$ conditionally independently
    from $\pdel(\cdot\mid x)$
    \State $A\gets b_\theta(x,u)-b_\theta(x,v)$
    \State Independently draw
    $H\sim\Geom(1-q)$
    \State $(\tilde x_0,\tilde u_0)\gets(x,u)$
    and $\widehat Q\gets0$
    \For{$h=0,\ldots,H$}
        \State Apply $\tilde u_h$ for duration $\delta$;
        observe $\tilde x_{h+1}$ and reward $r_h=r(\tilde x_h,\tilde u_h)$
        \State $\widehat Q
        \gets\widehat Q+q^h r_h$
        \If{$h<H$}
            \State Draw $\tilde u_{h+1}\sim
            \pdel(\cdot\mid\tilde x_{h+1})$
        \EndIf
    \EndFor
    \State \Return $\widehat Q A/(1-\gamma)$
\end{algorithmic}
\end{algorithm}
and provide the following results on the mean and variance of this estimator.
\begin{lemma}\label{lem:variance}
Suppose Assumptions~\ref{ass:safety} and~\ref{ass:policy} hold,
$\delta$ satisfies~\eqref{eq:period}, and sampling follows
Algorithm~\ref{alg:safegrad}. Then, with $V_\delta:=\frac{2M_{s,2}\bar r^2}
                    {(1-\gamma)^2(1-q)^2}$, we have
\begin{equation}\label{eq:V}
\begin{aligned}
 \E[\widehat g_\delta]&=\nabla J_\delta(\theta),\\
 \E\bigl[\norm{\widehat g_\delta-\nabla J_\delta(\theta)}^2\bigr]
 &\le V_\delta.
\end{aligned}
\end{equation}
Furthermore, each run of  Algorithm~\ref{alg:safegrad}  terminates almost surely after $(K+H+1)$ steps, with expected number of sampling intervals
\begin{equation}\label{eq:H}
 \bar{n}_{grad}:=\E[K+H+1]
 =\frac{\gamma}{1-\gamma}+\frac1{1-q}.
\end{equation}
\end{lemma}
\noindent\emph{Proof:} See Appendix~\ref{app:proofs}, item~(\ref{app:proof-estimator}).\par
With this estimator, we now design the following stochastic gradient ascent algorithm.
\subsection{SASPG algorithm}
We now describe the sampling-aware safe policy gradient method
\texttt{SASPG}, which updates the policy parameter $\theta$ to
maximize~\eqref{eq:intersample-safe-rl} over the class of $\pdel$.
Starting from $\theta_0$, at each iteration $j$ we hold
$\theta_j$ fixed and generate $B\ge1$ independent gradient
estimates using Algorithm~\ref{alg:safegrad}.
Then, we average these estimates and update the policy parameter
by stochastic gradient ascent with step size $\eta>0$. We summarize these steps in Algorithm~\ref{alg:SASPG}, and establish the convergence and optimality guarantees in the following section.
\begin{algorithm}[htbp]
\caption{\texttt{SASPG}: Sampling-Aware Safe Policy Gradient}
\label{alg:SASPG}
\begin{algorithmic}[1]
\Require $\theta_0\in\R^p$, $\gamma\in(0,1)$, $\delta$,
$\Cdel$, $\eta>0$,
$B\ge1$, $T\ge1$
\For{$j=0,\ldots,T-1$}
    \For{$b=1,\ldots,B$}
        \State $\widehat g_b
        \gets\texttt{SafeGrad}(\theta_j)$
    \EndFor
    \State $\displaystyle
    \overline g_j\gets\frac{1}{B}
    \sum_{b=1}^{B}\widehat g_b$
    \State $\theta_{j+1}
    \gets\theta_j+\eta\overline g_j$
\EndFor
\State \Return $\pi_{\theta_T}^\delta$
\end{algorithmic}
\end{algorithm}

\section{Convergence, Optimality, and Sample Complexity}
\label{sec:convergence}
We now establish finite-time convergence and sample-complexity
guarantees for \texttt{SASPG}. 
Here $L_\delta,V_\delta>0$ are upper bounds on the smoothness and
variance constants in~\eqref{eq:L} and~\eqref{eq:V}.
We now present the following assumption.
\begin{assumption}[Weak Gradient Dominance]\label{ass:gd}
There exist constants $a_\delta>0$ and $b_\delta\ge0$ such
that, for every $\theta\in\R^p$,
\begin{equation}
    J_\delta^*-J_\delta(\theta)\le
a_\delta\norm{\nabla J_\delta(\theta)}+b_\delta.
\end{equation}
\end{assumption}
This condition bounds the optimality gap by the gradient norm, up to
the residual $b_\delta$, and is used to characterize optimality
guarantees for nonconvex policy optimization
problems~\cite{montenegro2024learning,sharma2026safepg}.
Furthermore, such conditions are shown to hold for RL with classical control policies, e.g.
LQR \cite{fazel2018global,sharma2026safepg} and PID \cite{sharma2025globally}.

We now utilize results from Lemmas~\ref{lem:score}, \ref{lem:variance} on the smooth objective and an unbiased gradient estimator
with bounded variance to establish the convergence and sample-complexity
result for \texttt{SASPG} as follows.

\begin{theorem}[Finite-time Convergence]\label{thm:optimality}
Suppose Assumptions~\ref{ass:safety},~\ref{ass:policy}, and~\ref{ass:gd}
hold and $\delta$ satisfies~\eqref{eq:period}. Let
Algorithm~\ref{alg:SASPG} run from $\theta_0\in\R^p$ for
$T\ge1$ iterations with batch size $B\ge1$ and constant step
size $\eta$ satisfying
\begin{equation}\label{eq:stepsize}
\small
    0<\eta\le\eta_{\max}:=
 \min\left\{\frac{1}{L_\delta},
 \left(\frac{Ba_\delta^2}{L_\delta V_\delta}\right)^{1/3}\right\}.
\end{equation}
Then the following optimality distance holds
\begin{equation}\label{eq:gap}
\small
\begin{aligned}
 J_\delta^*-\E[J_\delta(\theta_T)]
 \le b_\delta+p_0\left(1-\sqrt{\frac{\eta^3L_\delta V_\delta}
                              {4a_\delta^2B}}\right)^T+a_\delta\sqrt{\frac{L_\delta V_\delta\eta}{B}},
\end{aligned}
\end{equation}
where $p_0:=\max\{0,J_\delta^*-J_\delta(\theta_0)-b_\delta\}$.
Furthermore, for any $\varepsilon>0$, and step size condition
\begin{equation}
\small
    \eta=\min\left\{\eta_{\max},
       \frac{B\varepsilon^2}{4a_\delta^2L_\delta V_\delta}\right\}.
\end{equation}
Then, to achieve $J_\delta^*-\E[J_\delta(\theta_T)]\le b_\delta+\varepsilon$, the number of iterations required follows
$T=\max\{1,\lceil\Lambda_\varepsilon/\lambda_\eta\rceil\}$ with the sample complexity following 
\begin{equation}
\small
    BT\le B+\Lambda_\varepsilon
\max\left\{
\frac{2a_\delta B^{3/2}L_\delta}{\sqrt{V_\delta}},\ 2B,\frac{16a_\delta^4L_\delta V_\delta}{\varepsilon^3}
\right\},
\end{equation}
where $\lambda_\eta:=\sqrt{\eta^3L_\delta V_\delta/(4a_\delta^2B)}$ and $\Lambda_\varepsilon:=\log\max\{1,2p_0/\varepsilon\}$.
\end{theorem}
\noindent\emph{Proof:} See Appendix~\ref{app:proofs}, item~(\ref{app:convergence-proof}).

Theorem~\ref{thm:optimality} gives sufficient conditions for
\texttt{SASPG} to reach a $(b_\delta+\varepsilon)$-neighborhood
of the optimal return within the safe policy class, with sample complexity scaling as
$O(\varepsilon^{-3}\log(1/\varepsilon))$. 
%


\section{Safe Quadrotor Experiment}
\label{sec:safe_quadrotor_experiment}
\color{black}
We evaluate \texttt{SASPG} on the quadrotor navigation setup
in~\cite{suttle2024sampling,sharma2026safepg}.
From the small-angle model
in~\cite{suttle2024sampling}, we consider the continuous-time quadrotor dynamics as
\begin{equation}\label{eq:quad-continuous}
 \dot p=v,\qquad \dot v=u,
\end{equation}
where $x=(p^\top,v^\top)^\top\in\R^4$ contains $x-y$ plane position
and velocity, and $u\in\U_{\mathrm{act}}=[-4,4]^2\,\mathrm{m/s^2}$
is the acceleration. The task is to reach
$p_{\mathrm g}=(15,15)\,\mathrm m$ from initial position
$p_0=(-1.5,-1.5)\,\mathrm m$ while avoiding an obstacle
centered at $(6.75,6.75)\,\mathrm m$. For safety, we model collision avoidance with $h(p)\ge0$, where CBF $h$ is given as
\begin{equation}\label{eq:quad_obstacle}
 h(p)=\left(\frac{p_x-6.75}{4}\right)^4
     +\left(\frac{p_y-6.75}{3}\right)^4-\tfrac12.
\end{equation}
We use the relative-degree-two barrier expression
$\psi=\ddot h+8\dot h+6h$~\cite{sharma2026safepg}.
Applying the margin construction~\eqref{eq:SASafeSet} with
$\underline\psi=\psi$ gives the sampling-aware safe control set
\begin{equation}\label{eq:quad-safe-set}
 \widehat{\mathcal C}_\delta(x)
 =\{u\in\U_{\mathrm{act}}:\psi(x,u)\ge\delta M_\psi(x,\delta)\},
\end{equation}
where $M_\psi(x,\delta)$ bounds $|\dot\psi|$ over a hold,
ensuring $\psi\ge0$ between updates. We choose
$\delta=0.10\,\mathrm s$ satisfying the criterion in
Proposition~\ref{prop:period},
$\delta M_\psi(x,\delta)<\max_{u\in\U_{\mathrm{act}}}\psi(x,u)$,
at each executed evaluation update. Under~\eqref{eq:zoh}, the sampled-data dynamics is given by
\begin{equation}\label{eq:quad-sampled}
 p_{k+1}=p_k+\delta v_k+\tfrac12\delta^2u_k,\quad
 v_{k+1}=v_k+\delta u_k.
\end{equation}
At sampling instant $k$, let $e_k=p_{\mathrm g}-p_k$ and
$\xi_k=(p_k,v_k,w_k)$. We use the Gaussian base policy
$\pi_\theta(\cdot\mid\xi_k)=\mathcal N(\mu_\theta(\xi_k),0.25I_2)$
centered at the PI controller~\cite{sharma2026safepg}
\begin{equation}\label{eq:quad-mean}
 \mu_\theta(\xi_k)=k_pe_k+k_i w_k-0.85v_k,
\end{equation}
where $\theta=(k_p,k_i)$ as PI parameters and $w_k\in\R^2$ as the accumulated error given by
\begin{equation}\label{eq:quad-memory}
 w_{k+1}=\operatorname{clip}
 \bigl(e^{-0.1 \delta}w_k+\delta e_k,-8,8\bigr),\quad w_0=0.
\end{equation}

\begin{figure}[!htb]
 \centering
 \includegraphics[width=\columnwidth,height=0.30\textheight,keepaspectratio]{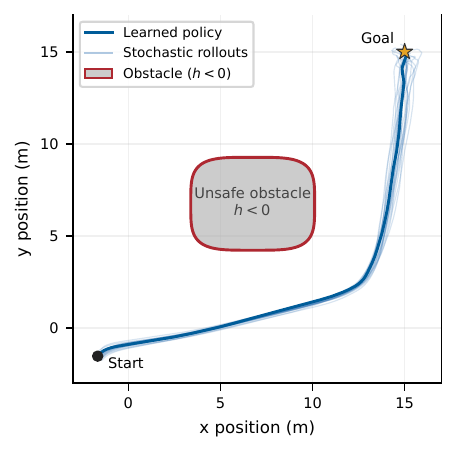}
 \caption{Quadrotor navigation using \texttt{SASPG}.
Light blue curves show 20 held-out trajectories with the
learned controller whereas the bold blue curve highlights its
median trajectory. The black dot and star
represent the quadrotor's initial and goal, respectively. The shaded
region represents the obstacle ($h<0$), with its boundary
($h=0$) shown in red.}
 \label{fig:quad_safe_navigation}
\end{figure}

For efficient sampling, we approximate $\widehat{\mathcal C}_\delta(x)$ with a maximum-area inner rectangle following~\cite{suttle2024sampling} and its
implementation~\cite{sharma2024cbfrepo}. Then, we obtain the safe-by-design policy $\pi_\theta^{\delta}$
from~\eqref{eq:policy}, and run
Algorithm~\ref{alg:safegrad} with return \eqref{eq:rl-return}, $\gamma=0.99$ and per-step reward
\begin{equation}
 r(p,u)=-\min\left\{\frac{\|p_{\mathrm g}-p\|_2}{20},1\right\}
        -\frac{0.01}{32}\|u\|_2^2.
\end{equation}
Using Algorithm~\ref{alg:SASPG}, we train ten policies for 500
updates each with batch size eight. We procide further implementation and evaluation details in Appendix~\ref{app:proofs}, item~(\ref{app:experiment-details}).
We show in Figure~\ref{fig:quad_safe_navigation} that \texttt{SASPG} learns a safe quadrotor navigation policy that is both safe at and between sampling instants. 

\section{Conclusion}\label{sec:conclusion}
We developed \texttt{SASPG}, a model-free policy gradient framework
for nonlinear sampled-data systems with hard CBF-based safety
constraints. We constructed stochastic policies supported on
sampling-period-dependent safe control sets and established
conditions ensuring constraint satisfaction throughout each
sampling interval. We provided finite-time convergence and
sample-complexity guarantees for the proposed algorithm and
characterized its optimality under weak gradient dominance.
We illustrated intersample safety on a quadrotor navigation task,
where the learned policy reaches the goal while avoiding an obstacle at all times.
Future work will explore alternative sampling and input-hold schemes,
including adaptive sampling through event-triggered updates, to
improve learning and control performance and reduce control updates
while preserving intersample safety.

\bibliographystyle{ieeetr}
\bibliography{references}

@inproceedings{achiam2017constrained,
 author={Achiam, Joshua and Held, David and Tamar, Aviv and Abbeel, Pieter},
 title={Constrained Policy Optimization},
 booktitle={Proc. International Conference on Machine Learning},
 series={Proceedings of Machine Learning Research}, volume={70},
 pages={22--31}, year={2017},
 url={https://proceedings.mlr.press/v70/achiam17a.html}}

@inproceedings{ames2019cbf,
 author={Ames, Aaron D. and Coogan, Samuel and Egerstedt, Magnus and
 Notomista, Gennaro and Sreenath, Koushil and Tabuada, Paulo},
 title={Control Barrier Functions: Theory and Applications},
 booktitle={Proc. European Control Conference}, pages={3420--3431},
 year={2019}, doi={10.23919/ECC.2019.8796030}}

@article{breeden2022sampled,
 author={Breeden, Joseph and Garg, Kunal and Panagou, Dimitra},
 title={Control Barrier Functions in Sampled-Data Systems},
 journal={IEEE Control Systems Letters}, volume={6}, pages={367--372},
 year={2022}, doi={10.1109/LCSYS.2021.3076127}}

@inproceedings{bahati2023samplehold,
 author={Bahati, Gilbert and Ong, Pio and Ames, Aaron D.},
 title={Sample-and-Hold Safety with Control Barrier Functions},
 booktitle={2024 American Control Conference (ACC)},
 pages={5169--5176}, year={2024},
 url={https://arxiv.org/abs/2304.08685}}

@inproceedings{suttle2024sampling,
 author={Suttle, Wesley and Sharma, Vipul Kumar and Kosaraju, Krishna Chaitanya
 and Seetharaman, Sivaranjani and Liu, Ji and Gupta, Vijay and Sadler, Brian M.},
 title={Sampling-based Safe Reinforcement Learning for Nonlinear Dynamical Systems},
 booktitle={Proc. International Conference on Artificial Intelligence and Statistics},
 series={Proceedings of Machine Learning Research}, volume={238},
 pages={4420--4428}, year={2024},
 url={https://proceedings.mlr.press/v238/suttle24a.html}}

@misc{sharma2026safepg,
 author={Sharma, Vipul K. and Suttle, Wesley A. and Sivaranjani, S.},
 title={{SafePG}: Safe and Globally Optimal Reinforcement Learning with Hard Constraints},
 year={2026}, note={arXiv:2609.14076},
 url={https://arxiv.org/abs/2609.14076}}

@article{xiao2022hocbf,
 author={Xiao, Wei and Belta, Calin},
 title={High-Order Control Barrier Functions},
 journal={IEEE Transactions on Automatic Control}, volume={67}, number={7},
 pages={3655--3662}, year={2022}, doi={10.1109/TAC.2021.3105491}}

@inproceedings{sutton1999policy,
 author={Sutton, Richard S. and McAllester, David and Singh, Satinder and Mansour, Yishay},
 title={Policy Gradient Methods for Reinforcement Learning with Function Approximation},
 booktitle={Advances in Neural Information Processing Systems},
 volume={12}, pages={1057--1063}, year={1999}}

@inproceedings{montenegro2024learning,
 author={Montenegro, Alessandro and Mussi, Marco and Metelli, Alberto Maria and Papini, Matteo},
 title={Learning Optimal Deterministic Policies with Stochastic Policy Gradients},
 booktitle={Proc. International Conference on Machine Learning},
 series={Proceedings of Machine Learning Research}, volume={235},
 pages={36160--36211}, year={2024},
 url={https://proceedings.mlr.press/v235/montenegro24a.html}}

@article{ames2017qp,
 author={Ames, Aaron D. and Xu, Xiangru and Grizzle, Jessy W. and Tabuada, Paulo},
 title={Control Barrier Function Based Quadratic Programs for Safety Critical Systems},
 journal={IEEE Transactions on Automatic Control}, volume={62}, number={8},
 pages={3861--3876}, year={2017}, doi={10.1109/TAC.2016.2638961}}

@article{hsu2024filter,
 author={Hsu, Kai-Chieh and Hu, Haimin and Fisac, Jaime F.},
 title={The Safety Filter: A Unified View of Safety-Critical Control in Autonomous Systems},
 journal={Annual Review of Control, Robotics, and Autonomous Systems},
 volume={7}, pages={47--72}, year={2024},
 doi={10.1146/annurev-control-071723-102940}}

@inproceedings{
sharma2025globally,
title={Globally Optimal Policy Gradient Algorithms for Reinforcement Learning with {PID} Control Policies},
author={Vipul Kumar Sharma and Wesley A. Suttle and S Sivaranjani},
booktitle={The Thirty-ninth Annual Conference on Neural Information Processing Systems},
year={2025},
url={https://openreview.net/forum?id=wW2qRfhT9d}
}

@misc{sharma2026safepglqr,
 author={Sharma, Vipul Kumar and Suttle, Wesley A. and Sivaranjani, S.},
 title={{Safe-PG-LQR}: Provably Safe, Convergent and Optimal Model-Free Linear Quadratic Regulator with Hard Constraints},
 howpublished={OpenReview},
 year={2026},
 note={\url{https://openreview.net/forum?id=iYjSjfIVcU}},
 url={https://openreview.net/forum?id=iYjSjfIVcU}
}

@misc{sharma2024cbfrepo,
  author       = {Sharma, Vipul Kumar},
  title        = {{CBF-constrained PPO}: Code for Sampling-based
                  Safe Reinforcement Learning},
  year         = {2024},
  howpublished = {GitHub repository.
                  \url{https://github.com/sharma1256/cbf-constrained_ppo}}
}

@inproceedings{fazel2018global,
 author={Fazel, Maryam and Ge, Rong and Kakade, Sham and Mesbahi, Mehran},
 title={Global Convergence of Policy Gradient Methods for the Linear Quadratic Regulator},
 booktitle={Proc. International Conference on Machine Learning},
 series={Proceedings of Machine Learning Research}, volume={80},
 pages={1467--1476}, year={2018},
 url={https://proceedings.mlr.press/v80/fazel18a.html}}

@inproceedings{yang2026cbf,
 author={Yang, Lizhi and Werner, Blake and de Sa, Massimiliano and Ames, Aaron D.},
 title={{CBF-RL}: Safety Filtering Reinforcement Learning in Training with Control Barrier Functions},
 booktitle={2026 IEEE International Conference on Robotics and Automation (ICRA)},
 pages={17825--17832}, year={2026},
 url={https://arxiv.org/abs/2510.14959}}

@inproceedings{singletary2020control,
  title={Control barrier functions for sampled-data systems with input delays},
  author={Singletary, Andrew and Chen, Yuxiao and Ames, Aaron D},
  booktitle={2020 59th IEEE Conference on Decision and Control (CDC)},
  pages={804--809},
  year={2020},
  organization={IEEE}
}

@inproceedings{cheng2019end,
  title={End-to-end safe reinforcement learning through barrier functions for safety-critical continuous control tasks},
  author={Cheng, Richard and Orosz, G{\'a}bor and Murray, Richard M and Burdick, Joel W},
  booktitle={Proceedings of the AAAI conference on artificial intelligence},
  volume={33},
  number={01},
  pages={3387--3395},
  year={2019}
}

@article{cohen2023safe,
 author={Cohen, Max H. and Belta, Calin},
 title={Safe Exploration in Model-based Reinforcement Learning using
 Control Barrier Functions},
 journal={Automatica}, volume={147}, pages={110684}, year={2023},
 doi={10.1016/j.automatica.2022.110684},
 url={https://arxiv.org/abs/2104.08171}}

@inproceedings{begzadic2026sampled,
 author={Begzadic, Azra and Lederer, Armin and Hirche, Sandra and
 Cort{\'e}s, Jorge and Herbert, Sylvia},
 title={Safe Event-Triggered Learning for Sampled-Data Systems},
 booktitle={2026 American Control Conference (ACC)},
 pages={805--810}, year={2026},
 url={https://portal.fis.tum.de/en/publications/safe-event-triggered-learning-for-sampled-data-systems/}}

@article{kochdumper2023projection,
 author={Kochdumper, Niklas and Krasowski, Hanna and Wang, Xiao and
 Bak, Stanley and Althoff, Matthias},
 title={Provably Safe Reinforcement Learning via Action Projection Using Reachability Analysis and Polynomial Zonotopes},
 journal={IEEE Open Journal of Control Systems},
 volume={2}, pages={79--92}, year={2023},
 doi={10.1109/OJCSYS.2023.3256305},
 url={https://arxiv.org/abs/2210.10691}}

@article{lanza2024sampled,
 author={Lanza, Lukas and Dennst{\"a}dt, Dario and Worthmann, Karl and
 Schmitz, Philipp and {\c{S}}en, G{\"o}k{\c{c}}en Devlet and Trenn, Stephan
 and Schaller, Manuel},
 title={Sampled-data funnel control and its use for safe continual learning},
 journal={Systems \& Control Letters}, volume={192}, pages={105892},
 year={2024}, doi={10.1016/j.sysconle.2024.105892},
 url={https://arxiv.org/abs/2303.00523}}
\useRomanappendicesfalse
\appendices
\section{Proofs and Experimental Details}

\label{app:proofs}

\begin{enumerate}[label=(\roman*),ref=\roman*,leftmargin=0pt,itemindent=2em,labelwidth=1.5em,labelsep=0.5em,itemsep=6pt,topsep=0pt]

\item\label{app:proof-safety}\textbf{Proof of Theorem~\ref{thm:safety}.}
For $x_k\in\Ssafe$ and $u_k\in\Cdel(x_k)$,
Assumption~\ref{ass:safety} and~\eqref{eq:SASafeSet} imply, for
$0\le s\le\delta$,
\begin{equation}\label{eq:lower}
\begin{aligned}
 \psi(x(k\delta+s),u_k)
 &\ge\underline\psi(x_k,u_k)-M_\psi s\\
 &\ge M_\psi(\delta-s)\ge0.
\end{aligned}
\end{equation}
Hence $\dot h(x(t))+\alpha_1(h(x(t)))\ge0$ almost everywhere on
$[k\delta,(k+1)\delta]$. Scalar comparison preserves $h(x(t))\ge0$
from $h(x_k)\ge0$. Induction over the sampling intervals proves
forward invariance of $\Ssafe$, and the uniform existence condition
permits continuation at each update. The actuator constraints follow
from $\Cdel(x_k)\subseteq\U_{\mathrm{act}}$.\hfill\(\square\)\par

\item\label{app:proof-period}\textbf{Proof of Proposition~\ref{prop:period}.}
We show $\lambda_m(\Cdel(x))>0$ for every $x\in\Ssafe$.
By definition, $m_{\mathrm{act}}=\inf_{x\in\Ssafe}
\max_{u\in\U_{\mathrm{act}}}\underline\psi(x,u)$.
For each $x\in\Ssafe$, continuity of $\underline\psi(x,\cdot)$ and
compactness of $\U_{\mathrm{act}}$ give a maximizing input $u^\star$
with $\underline\psi(x,u^\star)\ge m_{\mathrm{act}}>M_\psi\delta$.
Continuity preserves this strict inequality in a neighborhood of
$u^\star$ within the actuator box. Since the box has nonempty interior,
this neighborhood has positive volume and belongs to $\Cdel(x)$.\hfill\(\square\)\par

\item\label{app:proof-smoothness}\textbf{Proof of Lemma~\ref{lem:score}.}
Assumption~\ref{ass:policy}(\ref{asum-item-integrability})
justifies differentiation of the normalizer and finite-horizon
trajectory expectations. Since $\Cdel(x)$ is
independent of $\theta$, differentiating the normalizer gives
$\nabla_\theta\log Z_\theta^\delta(x)
=\E_{v\sim\pdel(\cdot\mid x)}[b_\theta(x,v)]$,
which yields~\eqref{eq:score}.
Set $G:=\sqrt{2M_{s,2}}$ and $H_s:=M_{D_s}$.
For a reward $r_k=r(x_k,u_k)$, write
$S_k=\sum_{i=0}^k s_\theta^\delta(x_i,u_i)$ and
$T_k=\sum_{i=0}^kD_\theta s_\theta^\delta(x_i,u_i)$.
Finite-horizon likelihood differentiation gives
$\nabla_\theta^2\E_\theta^\delta[r_k]
=\E_\theta^\delta[r_k(S_kS_k^\top+T_k)]$.
The moment conditions imply
$\E_\theta^\delta[\norm{S_k}^2]\le(k+1)^2G^2$ and
$\E_\theta^\delta[\norm{T_k}]\le(k+1)H_s$. Hence
\[
 \norm{\nabla_\theta^2\E_\theta^\delta[r_k]}
 \le\bar r[(k+1)^2G^2+(k+1)H_s].
\]
Also,
$\norm{\nabla_\theta\E_\theta^\delta[r_k]}
\le\bar r(k+1)\sqrt{M_{s,2}}$.
After multiplication by $\gamma^k$, both bounds are summable uniformly
in $\theta$, justifying differentiation of the infinite-horizon return.
Using
$\sum_{k\ge0}\gamma^k(k+1)^2=(1+\gamma)/(1-\gamma)^3$
and $\sum_{k\ge0}\gamma^k(k+1)=(1-\gamma)^{-2}$
in the Hessian bound yields~\eqref{eq:L}.
The gradient inequality follows by integration along the segment
joining $\theta$ and $\theta'$.\hfill\(\square\)\par

\item\label{app:proof-estimator}\textbf{Proof of Lemma~\ref{lem:variance}.}
Fix $\theta$ and let $K\sim\Geom(1-\gamma)$ be independent
of the reset state and rollout. Then
\[
 \mathcal L_\theta^\delta(x_K)
 =(1-\gamma)\sum_{k=0}^{\infty}
       \gamma^k\mathcal L_\theta^\delta(x_k)
 =d_\theta^\delta.
\]
At $x=x_K$, draw $u,v$ conditionally independently from
$\pdel(\cdot\mid x)$ and set
$A_\theta(x,u,v)=b_\theta(x,u)-b_\theta(x,v)$.
By~\eqref{eq:score},
\[
 \E[A_\theta(x,u,v)\mid x,u]=s_\theta^\delta(x,u).
\]
With $q=\sqrt\gamma$ and an independent $H\sim\Geom(1-q)$,
the randomized return satisfies
\begin{equation}\label{eq:Qhat}
 \widehat Q_\delta=\sum_{h=0}^H q^h r_h,
 \qquad |\widehat Q_\delta|\le\frac{\bar r}{1-q}.
\end{equation}
Since $\Pr(H\ge h)=q^h$ and $q^2=\gamma$, bounded rewards
allow interchange of expectation and summation, giving
$\E[\widehat Q_\delta\mid x,u]=Q_\theta^\delta(x,u)$.
Conditional independence of $v$ and the return continuation then gives
\[
 \E[\widehat Q_\delta A_\theta(x,u,v)\mid x,u]
 =Q_\theta^\delta(x,u)s_\theta^\delta(x,u).
\]
Equation~\eqref{eq:pg} proves unbiasedness.
The paired-score second moment is
\begin{equation}\label{eq:pair-moment}
\begin{aligned}
 \E[\norm{A_\theta(x,u,v)}^2\mid x]
 &=2\E[\norm{s_\theta^\delta(x,u)}^2\mid x]\\
 &\le2M_{s,2}.
\end{aligned}
\end{equation}
Together with~\eqref{eq:Qhat}, this gives
$\E[\norm{\widehat g_\delta}^2]\le V_\delta$, and hence the variance bound.

The first stage applies $K$ inputs, and the second applies
$H+1$ inputs, including the initial action $u$.
The auxiliary action $v$ is not applied to the plant.
Since $0<\gamma,q<1$, both stopping times are finite almost surely.
Their geometric distributions give
$\E[K]=\gamma/(1-\gamma)$ and $\E[H+1]=1/(1-q)$,
which prove~\eqref{eq:H}. For a fixed number $N_{\rm PG}$ of
estimator calls, linearity of expectation gives
$N_{\rm PG}\bar{n}_{grad}$ actuated intervals in expectation,
including when the policy parameter changes between calls.\hfill\(\square\)\par

\item\label{app:convergence-proof}\textbf{Proof of Theorem~\ref{thm:optimality}.}
Let $\mathcal F_j$ contain all randomness before batch $j$, so
$\theta_j$ is $\mathcal F_j$-measurable. Independent calls to
\textsc{SafeGrad} and Lemma~\ref{lem:variance} give
\begin{equation}\label{eq:batch}
\begin{aligned}
 \E[\overline g_j\mid\mathcal F_j]&=\nabla J_\delta(\theta_j),\\
 \E[\norm{\overline g_j-\nabla J_\delta(\theta_j)}^2
       \mid\mathcal F_j]&\le V_\delta/B.
\end{aligned}
\end{equation}
For $\eta\le L_\delta^{-1}$, smoothness and~\eqref{eq:batch} yield
\begin{equation}\label{eq:ascent}
\begin{aligned}
 \E[J_\delta(\theta_{j+1})\mid\mathcal F_j]
 &\ge J_\delta(\theta_j)
   +\frac\eta2\norm{\nabla J_\delta(\theta_j)}^2\\
 &\quad-\frac{L_\delta\eta^2V_\delta}{2B}.
\end{aligned}
\end{equation}
Define $z_j=J_\delta^*-\E[J_\delta(\theta_j)]-b_\delta$,
$c=\eta/(2a_\delta^2)$, and $d=L_\delta\eta^2V_\delta/(2B)$.
Write $[z]_+=\max\{z,0\}$.
Weak gradient dominance and Jensen's inequality imply
$\E[\norm{\nabla J_\delta(\theta_j)}^2]
\ge[z_j]_+^2/a_\delta^2$, hence
\begin{equation*}
 z_{j+1}\le z_j-c[z_j]_+^2+d.
\end{equation*}
Write $p_0=[z_0]_+$ and set
\begin{equation}\label{eq:lambda}
\begin{aligned}
 r=\xi_\eta&=a_\delta\sqrt{\frac{L_\delta\eta V_\delta}{B}},\\
 \lambda=\lambda_\eta&=cr=\frac{\eta\xi_\eta}{2a_\delta^2}.
\end{aligned}
\end{equation}
The step-size restriction~\eqref{eq:stepsize} gives
$0<\lambda\le1/2$.
For $z_j\le r$, the recursion gives $z_{j+1}\le r$:
for $z_j\le0$ use $d=\lambda r\le r$, and on $[0,r]$
use monotonicity of $z-cz^2+d$. For $z_j>r$,
\begin{equation*}
 z_j-cz_j^2+d
 =r+(1-2\lambda)(z_j-r)-c(z_j-r)^2.
\end{equation*}
Thus $[z_{j+1}-r]_+\le(1-\lambda)[z_j-r]_+$.
Iteration and $[z_0-r]_+\le p_0$ prove~\eqref{eq:gap}.

For any $\varepsilon>0$, choose
\begin{equation}\label{eq:eta}
 \eta=\min\left\{\eta_{\max},
       \frac{B\varepsilon^2}{4a_\delta^2L_\delta V_\delta}\right\},
\end{equation}
so $r\le\varepsilon/2$. Define
$\Lambda_\varepsilon=\log\max\{1,2p_0/\varepsilon\}$ and set
\begin{equation}\label{eq:iterations}
 T=\max\left\{1,
      \left\lceil\frac{\Lambda_\varepsilon}{\lambda}\right\rceil
      \right\}.
\end{equation}
Since $(1-\lambda)^T p_0\le\varepsilon/2$,
\eqref{eq:gap} gives $b_\delta+\varepsilon$ accuracy.
Also, $N_{\rm PG}=BT\le B+B\Lambda_\varepsilon/\lambda$;
substitution of~\eqref{eq:eta} gives
\begin{equation}\label{eq:complexity}
\begin{aligned}
 N_{\rm PG}\le B+\Lambda_\varepsilon\max\Biggl\{
 &\frac{2a_\delta B^{3/2}L_\delta}{\sqrt{V_\delta}},\ 2B,\\
 &\frac{16a_\delta^4L_\delta V_\delta}{\varepsilon^3}\Biggr\}.
\end{aligned}
\end{equation}
This yields the claimed small-$\varepsilon$ order for fixed $B$ and problem constants.
Equation~\eqref{eq:H} gives the expected interaction count.
\hfill\(\square\)\par

\end{enumerate}

\begin{remark}[Stationarity without weak gradient dominance]
\label{rem:stationarity}
Under Assumptions~\ref{ass:safety} and~\ref{ass:policy} and the
period condition~\eqref{eq:period}, telescoping~\eqref{eq:ascent}
yields, for $0<\eta\le L_\delta^{-1}$ and integer $T\ge1$,
\begin{equation}\label{eq:stationary}
\begin{aligned}
 \frac1T\sum_{j=0}^{T-1}\E[\norm{\nabla J_\delta(\theta_j)}^2]
 &\le\frac{2[J_\delta^*-J_\delta(\theta_0)]}{\eta T}\\
 &\quad+\frac{L_\delta\eta V_\delta}{B}.
\end{aligned}
\end{equation}
This uses smoothness,~\eqref{eq:batch}, and
$\E[J_\delta(\theta_T)]\le J_\delta^*$.
\end{remark}
\begin{enumerate}[label=(\roman*),ref=\roman*,start=6,leftmargin=0pt,itemindent=2em,labelwidth=1.5em,labelsep=0.5em,itemsep=6pt,topsep=0pt]
\item\label{app:experiment-details}\textbf{Experimental details.}\par
\noindent\textit{Initial Variables:} 
We draw $p_0$ uniformly from $(-1.5,-1.5)+[-0.25,0.25]^2$
and $v_0$ from $[-0.1,0.1]^2$, with $w_0=0$. Initial states satisfy $h\ge0$ and
$\dot h+(4-\sqrt{10})h\ge0$.

\noindent\textit{Barrier Margin and Inner Rectangle:} 
Suppose $\rho_i=p_i-o_i$, with $o=(6.75,6.75)$ and
$(a_1,a_2)=(4,3)$. Further, suppose $U=4$, $Q_i=|\rho_i|+\delta|v_i|+\tfrac12\delta^2U$ and
$V_i=|v_i|+\delta U$, which bound position and velocity.
Then, differentiating $\psi=\ddot h+8\dot h+6h$, we get
\begin{equation}\label{eq:quad-M}
 \begin{aligned}
 M_\psi(x,\delta)=\sum_{i=1}^2 &\frac1{a_i^4}\Bigl[
 24Q_iV_i^3+36Q_i^2V_iU \\
 &+8(12Q_i^2V_i^2+4Q_i^3U)+24Q_i^3V_i\Bigr].
 \end{aligned}
\end{equation}
Thus $\psi(x(t_k+s),u_k)\ge\psi(x_k,u_k)-M_\psi(x_k,\delta)s$.
The affine input constraint and its maximum-area inner rectangle
follow~\cite[Appendices C.2.2--C.2.3]{suttle2024sampling}
and~\cite{sharma2024cbfrepo}, with the threshold tightened by
$\delta M_\psi$ as in~\eqref{eq:quad-safe-set}.

\noindent\textit{Gaussian Truncation and Gradient Estimates:} 
We sample the rectangle-truncated Gaussian coordinatewise by
inverse-CDF sampling, with $\sigma=0.5$.
The paired score follows from~\eqref{eq:score} using
$b_\theta=\sigma^{-2}(D_\theta\mu_\theta)^\top(u-\mu_\theta)$
and $D_\theta\mu_\theta=[p_{\mathrm g}-p\ \ w]$.
Algorithm~\ref{alg:safegrad} supplies the gradient estimates.
We initialize $(k_p,k_i)=(0.18,0)$ and use step size as $10^{-4}$.

\noindent\textit{Termination and Evaluation:} 
Each controller is tested on 100 common held-out training runs for
$40\,\mathrm s$. 
After goal completion, the training enters an absorbing terminal
state with zero reward at each subsequent step. If an infeasible
safe control set, sampling failure, collision, or escape beyond
the prescribed position boundaries occurs, the training instead enters
an absorbing terminal state with a reward $-1.01$ at each subsequent
step.
Figure~\ref{fig:quad_safe_navigation} uses the controller with
lower-median mean evaluation return across the ten training runs.
\end{enumerate}

\end{document}